\documentclass[runningheads]{llncs}

\usepackage[T1]{fontenc}
\usepackage{graphicx}
\usepackage{booktabs}
\usepackage[misc]{ifsym}
\usepackage{subfigure}
\usepackage{bm}
\usepackage{stmaryrd}
\usepackage{balance}
\usepackage{pgfplots}
\usepackage{pifont}
\usepackage{float}
\usepackage{epsfig}
\usepackage{pgffor}
\usepackage{tikz}
\usepackage{bbm}
\usepackage{helvet}
\usepackage{courier}
\usepackage{threeparttable}
\usepackage{algpseudocode} 
\usepackage{arydshln}
\usepackage{mathrsfs}
\usepackage{rotating} 
\usepackage{adjustbox}
\usepackage{makecell}
\usepackage{diagbox} 
\usepackage[skip=0pt]{caption}
\usepackage{subfloat}
\usepackage{amssymb}
\usepackage{amsfonts}
\usepackage{amsmath}
\usepackage{enumitem}

\newcommand{\corr}{(\Letter)}
\usepackage{mwe}

\begin{document}

\title{Hyperbolic Contrastive Learning with Model-augmentation for Knowledge-aware Recommendation}

\titlerunning{Hyperbolic Model-contrastive for Knowledge-aware Recommendation}

\author{Shengyin Sun \and
Chen Ma \corr}

\authorrunning{S. Sun et al.}

\institute{City University of Hong Kong, 83 Tat Chee Avenue, Kowloon, Hong Kong, China \email{\{shengysun4-c@my.cityu.edu.hk, chenma@cityu.edu.hk\}}
}

\tocauthor{Shengyin Sun, Chen Ma}
\toctitle{Hyperbolic Contrastive Learning with Model-augmentation for Knowledge-aware Recommendation}

\maketitle              

\begin{abstract}
Benefiting from the effectiveness of graph neural networks (GNNs) and contrastive learning, GNN-based contrastive learning has become mainstream for knowledge-aware recommendation. However, most existing contrastive learning-based methods have difficulties in effectively capturing the underlying hierarchical structure within user-item bipartite graphs and knowledge graphs. Moreover, they commonly generate positive samples for contrastive learning by perturbing the graph structure, which may lead to a shift in user preference learning. To overcome these limitations, we propose hyperbolic contrastive learning with model-augmentation for knowledge-aware recommendation. To capture the intrinsic hierarchical graph structures, we first design a novel Lorentzian knowledge aggregation mechanism, which enables more effective representations of users and items. Then, we propose three model-level augmentation techniques to assist Hyperbolic contrastive learning. Different from the classical structure-level augmentation (e.g., edge dropping), the proposed model-augmentations can avoid preference shifts between the augmented positive pair. Finally, we conduct extensive experiments to demonstrate the superiority (maximum improvement of $11.03\%$) of proposed methods over existing baselines.\footnote{\url{Code available at https://github.com/sunshy-1/HCMKR}.}

\keywords{Knowledge-aware recommendation \and Model-augmentation \and Hyperbolic space.}
\end{abstract}

\section{Introduction}
\label{ssy1210:introduction}
Recommender systems have been widely used in many real-world online services (e.g., online shopping and online advertising) to perform personalized information filtering. With the rapid development of neural networks, the classical collaborative filtering paradigm in recommender systems has evolved from matrix factorization to learning user/item representations with neural networks~\cite{DBLP:journals/RS/GaoZheng23}. Considering that user-item historical records can be regarded as a user-item bipartite graph, many studies~\cite{DBLP:conf/sigir/HeDeng20,DBLP:journals/IS/LiYang24} have recently focused on applying graph neural networks (GNNs) in the collaborative filtering paradigm. The core idea of GNN-based methods is to learn user/item representations through multiple hops of neighborhood aggregation. The neighbor aggregation mechanism is expected to capture the high-order connectivity within user-item interactions, thereby improving the quality of representations.

\begin{figure}
    \centering
    \subfigure[Amazon-Book]{
        \includegraphics[width=2.2in]{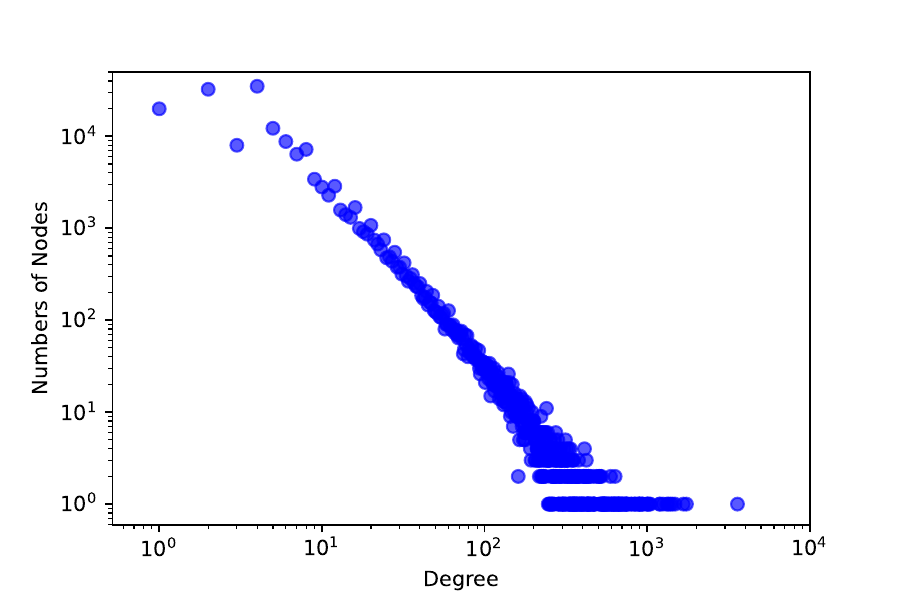}}
    \subfigure[MovieLens-20M]{
	\includegraphics[width=2.2in]{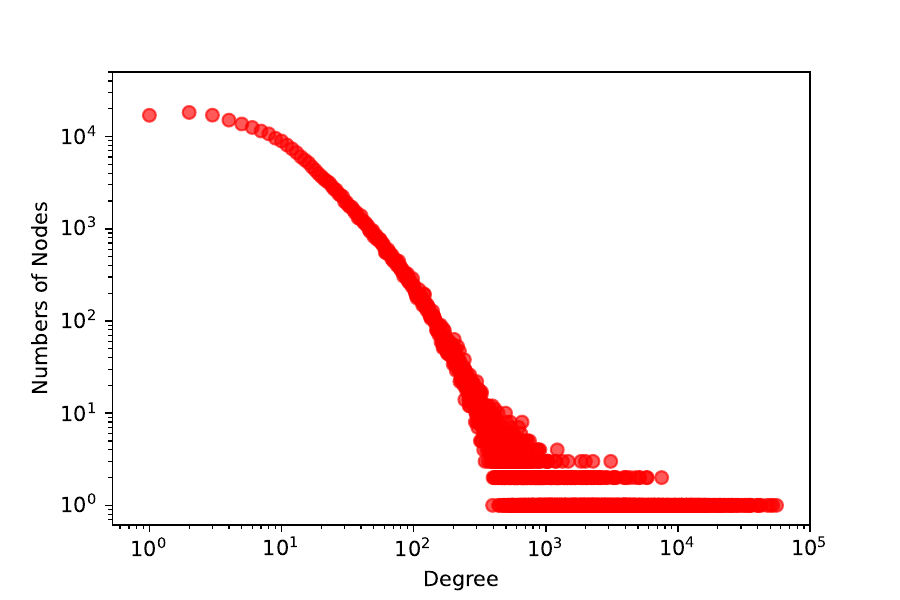}}
    \caption{Degree distributions of two public datasets.}
    \label{ssy0507:degree_distribution}
\end{figure}

To further improve the performance and explainability of the model, knowledge-aware recommender systems came into being. They integrate side information--knowledge graphs into traditional collaborative filtering models. Motivated by the capability of GNNs and contrastive learning in various fields, designing GNN-based knowledge-aware methods using contrastive learning has recently become promising~\cite{DBLP:conf/sigir/ZouWei22,DBLP:conf/sigir/YangHuang22}. Although existing contrastive learning-based knowledge-aware methods have made substantial progress, we argue that most of them have the following limitations:
\begin{itemize}
    \item[$\bullet$] {\textbf{Ineffective to capture hierarchical structures in user-item bipartite graphs and knowledge graphs.} The degree distributions of two public datasets are shown in Fig.~\ref{ssy0507:degree_distribution}. It is obvious that the distributions in Fig.~\ref{ssy0507:degree_distribution} approximate the power-law distribution. According to Bourgain's theorem~\cite{DBLP:journals/Combinatorica/LinialLondon95}, the Euclidean space lacks the ability to obtain comparably low distortion for the power-law distributed graph data~\cite{DBLP:conf/icml/SalaSa18}. Thus, conventional methods~\cite{DBLP:conf/sigir/YangHuang22,DBLP:conf/sigir/ZouWei22} that perform representation learning in the Euclidean space cannot effectively capture the underlying hierarchical structure contained in the power-law distributed graph data~\cite{DBLP:conf/www/SunCheng21,DBLP:conf/wsdm/ChenYang22}.}
    \item[$\bullet$] {\textbf{Inadequate to maintain user interest context and avoid preference shift.} Existing contrastive learning methods~\cite{DBLP:conf/sigir/WuWang21,DBLP:conf/sigir/YangHuang22} generate positive pairs by perturbing the graph structure. However, in recommendation scenarios, the original graph structure serves as the supervision signal for learning user preferences. By perturbing the graph structure, the items that are used to learn the user preference will be altered, which potentially shifts the user interest and jeopardizes the stability of learned user preferences.}
\end{itemize}

\begin{figure}
    \centering
    \includegraphics[scale=0.38]{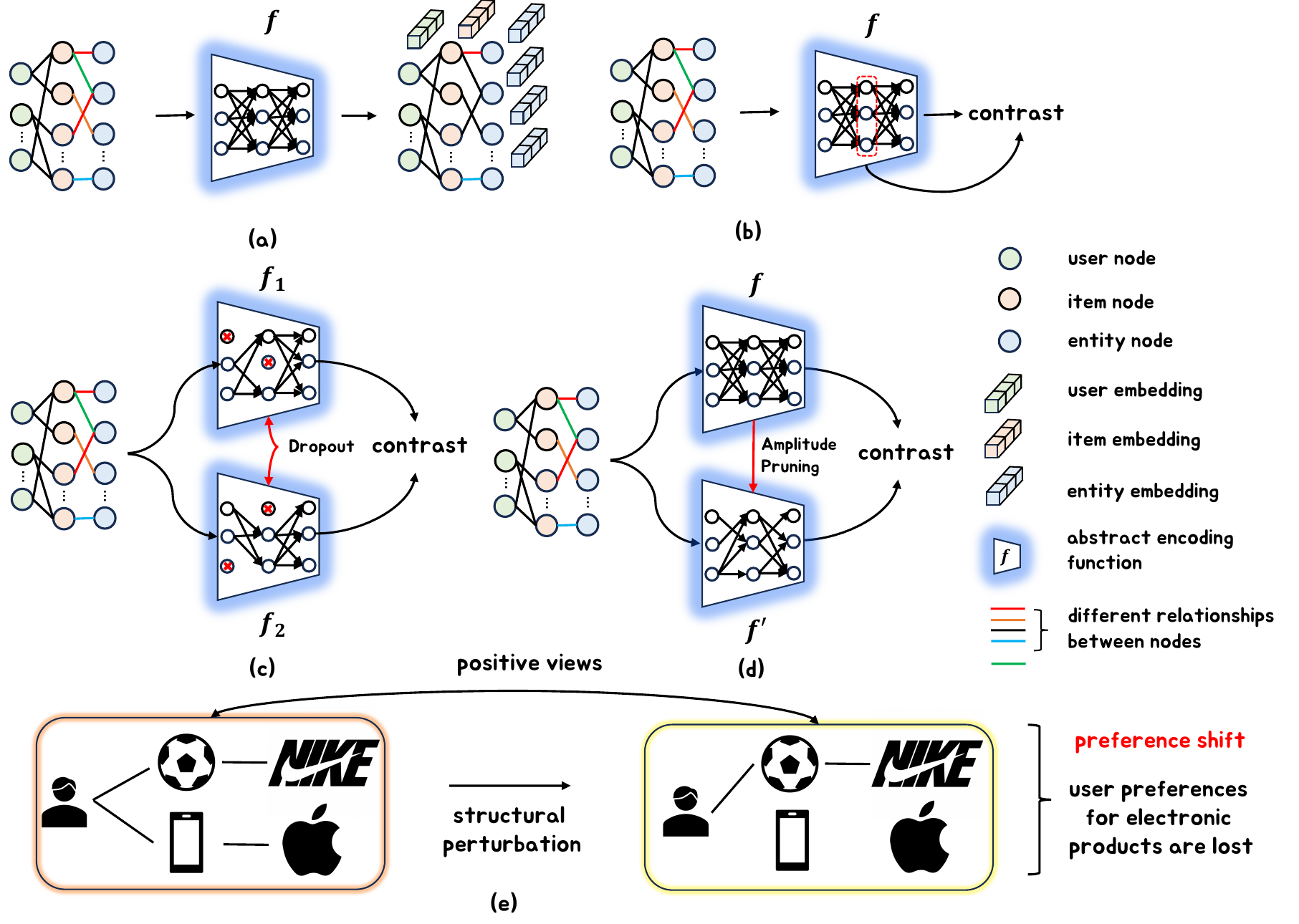}
    \caption{Illustrations of the model-level augmentations and preference shift. (a) An abstract process of representation learning in the GNN-based knowledge-aware recommender systems, where $f(\cdot)$ is the abstract encoding function. We mainly focus on perturbing the structure of $f(\cdot)$ to generate augmented samples. (b) Augmentations via cross-layer outputs. (c) Augmentations via Dropout. (d) Augmentations via model-pruning. (e) The structural perturbation leads to the loss of the user preference for electronic products.}
\label{ssy0507:illustration_for_model_level_aug}
\end{figure}

In order to simultaneously overcome the above two limitations, we propose the \emph{\underline{H}yperbolic \underline{C}ontrastive Learning with \underline{M}odel-augmentation for \underline{K}nowledge-aware \underline{R}ecommendation} (HCMKR). To effectively capture the intrinsic hierarchical structures, we design a framework for representation learning based on the Lorentz model (as detailed in Section~\ref{ssy0507:hyperbolic_geometry}). Specifically, we first design a Lorentzian knowledge aggregation mechanism, and then encode the representations of users/items in the Lorentz-based Hyperbolic space. To address the issue of preference shift, we propose model-level augmentation techniques to introduce auxiliary supervision signals, i.e., augmentations via Dropout, cross-layer outputs (which can be regarded as the model layer dropping), and model pruning. The proposed model-augmentation focuses on the perturbation at the model level (as shown in Fig.~\ref{ssy0507:illustration_for_model_level_aug}), which is distinct from the classical augmentation techniques that directly perturb graph structures~\cite{DBLP:conf/sigir/YangHuang22,DBLP:conf/sigir/ZouWei22}. Directly perturbing the graph structure may cause preference shifts and semantic changes. As shown in Fig.~\ref{ssy0507:illustration_for_model_level_aug} (e), deleting edges on the user-item graph is equivalent to modifying the user’s purchase history to a certain extent, leading to changes in the semantics of user representations. On the contrary, the proposed model augmentation has a natural advantage over classical augmentation by keeping user interest context and avoiding preference shift. 

The main contributions of this paper are summarized below:
\begin{itemize}
    \item[$\bullet$] {To capture the hierarchical characteristics of graphs, we propose Hyperbolic contrastive learning for the knowledge-aware recommendation. Specifically, we design a novel Lorentzian knowledge aggregation mechanism, which enables effective modeling of user/item representations. }
    \item[$\bullet$] {We introduce model-augmentation techniques into the contrastive learning-based knowledge-aware recommendation. More specifically, we propose three model-level augmentation techniques to offer additional supervision signals. The proposed model-level augmentation does not perturb the graph structure, which enables better semantic preservation and avoids preference shifts between the augmented positive pair.}
    \item[$\bullet$] {Extensive experiments are conducted on three datasets to assess the performance of the proposed methods. Experimental results show the effectiveness of the proposed methods. Compared with the state-of-the-art baselines, the proposed methods achieve a maximum performance improvement of 11\% with reduced time consumption ($\times 0.88$).}
\end{itemize}
\section{Preliminary}
\subsection{Notations}
 Matrices and vectors are denoted by bold upper case letters (e.g., $\mathbf{{A}}$) and bold lower case letters (e.g., $\mathbf{{a}}$), respectively. Sets are denoted by calligraphic letters (e.g., $\mathcal{U}$). $\vert \cdot \vert$ represents the number of elements in the set (e.g., $\vert\mathcal{U}\vert$). Superscript $(\cdot)^{\top}$ denotes transpose. $\|$ is the concatenation operation.

\subsection{User-Item Bipartite Graph}
A user-item bipartite graph is a graph that contains two types of nodes (user and item) and one type of edges (user-item interaction). Define a user-item bipartite graph as $\mathcal{G}_{1}=\{\mathcal{U}, \mathcal{I}, \mathcal{E}, \mathbf{{A}}\}$, where $\mathcal{U}$ is the set of user nodes, $\mathcal{I}$ is the set of item nodes, $\mathcal{E}$ is the set of user-item interaction edges, and $\mathbf{{A}}$ is an adjacency matrix. If there is an interaction between the user $u$ and the item $i$, then there is an edge $(u,i)$ in $\mathcal{E}$ and $\mathbf{A}_{ui}=1$.

\subsection{Knowledge Graph}
In addition to user-item interaction data, we also have side information for items. The side information contains rich semantic information that facilitates user preference modeling and is usually organized in the form of a knowledge graph~\cite{DBLP:conf/kdd/WangHe19}. Formally, a knowledge graph $\mathcal{G}_2$ is defined as a set of entity-relation-entity triplets, i.e., $\mathcal{G}_2=\{(h,r,t), h,t\in\mathcal{E}^{\prime}, r\in\mathcal{R}^{\prime}\}$, where $h$ is a head entity, $t$ is a tail entity, $r$ represents the relationship between $h$ and $t$, $\mathcal{E}^{\prime}$ is the set of entities, and $\mathcal{R}^{\prime}$ is the set of relations. With the aid of the knowledge graph, the task in our hands can be formulated as follows: Given a user-item bipartite graph $\mathcal{G}_1$ and an item knowledge graph $\mathcal{G}_2$, train a model $f(u,i|\mathcal{G}_1, \mathcal{G}_2, \boldsymbol{\Theta})$ to predict the probability of user $u$ interacting with item $i$, where $u\in\mathcal{U}$, $i\in\mathcal{I}$, and $\boldsymbol{\Theta}$ is a matrix of trainable parameters.

\section{Methodology}
\label{ssy1210:method}
\subsection{Hyperbolic Geometry}
\label{ssy0507:hyperbolic_geometry}
As for Hyperbolic geometry, multiple Hyperbolic models have been proposed, such as the Poincaré ball model and the Lorentz model. Among them, we use the Lorentz model due to its superiority in terms of numerical stability {\textcolor{black}{(the distance function in the Lorentz model avoids numerical instabilities that arise from the fraction in the Poincaré distance)}}~\cite{DBLP:conf/icml/LawLiao19,DBLP:conf/nips/ChamiYing19}.

\vspace{5pt}
\noindent\textbf{$\bullet$ Lorentz model} Denote the Lorentz model in $d$-dimensional space with the negative curvature $-\frac{1}{c}$ ($c>0$) by $\mathbb{L}^{d,c}$. The Lorentz model $\mathbb{L}^{d,c}$ is defined as a subset of $\mathbb{R}^{d}$, i.e.,
\begin{equation}
    \label{ssy0507:LorentzModel}
    \mathbb{L}^{d, c}=\left\{\mathbf{x}=\left[x_0,x_1,\cdots,x_{d-1}\right] \in \mathbb{R}^{d} \mid\langle\mathbf{x}, \mathbf{x}\rangle_{\mathbb{L}}=-c\right\},
\end{equation}
where $\langle\cdot, \cdot\rangle_{\mathbb{L}}:\mathbb{R}^{d}\times\mathbb{R}^{d}\rightarrow\mathbb{R}$ is the Lorentz inner product. Let $\mathbf{a},\mathbf{b}\in\mathbb{R}^{d}$, then the Lorentz inner product can be written as: $\langle\mathbf{a}, \mathbf{b}\rangle_{\mathbb{L}} = -a_0 b_0+\sum_{i=1}^{d-1} a_i b_i.$

\vspace{5pt}
\noindent\textbf{$\bullet$ Tangent space} The tangent space is a first-order approximation of the Lorentzian manifold at a specific point, which is widely used in hyperbolic graph convolutional networks. Mathematically, the tangent space $\mathbb{T}_{\mathbf{x}}^{d,c}$ with the origin $\mathbf{x}\in\mathbb{L}^{d,c}$ (the reference point) is defined as follows:
\begin{equation}
    \label{ssy0507:TanSpace}
    \mathbb{T}_{\mathbf{x}}^{d,c} = \left\{\mathbf{x^{\prime}}=\left[x_0^{\prime},x_1^{\prime},\cdots,x_{d-1}^{\prime}\right] \in \mathbb{R}^{d} \mid\langle\mathbf{x^{\prime}}, \mathbf{x}\rangle_{\mathbb{L}}=0\right\}.
\end{equation}

\subsection{Lorentzian Knowledge Aggregation}
\label{ssy0507:KnowAgg}
{\textbf{$\bullet$ Space Mapping}}
Considering that the input embedding is usually in the Euclidean space, we first map the Euclidean embeddings to the Hyperbolic embeddings (the so-called space mapping). Two kinds of mapping functions are frequently used in space mapping, namely the logarithmic and the exponential mapping functions. Mathematically, given $\mathbf{t},\mathbf{u}\in\mathbb{L}^{d,c}~(\mathbf{t} \neq \mathbf{u})$ and $\mathbf{v}\in\mathbb{T}_{\mathbf{t}}^{d,c}$, the logarithmic mapping function $\Pi^{log,c}_{\mathbf{t}}(\mathbf{u}):\mathbb{L}^{d,c}\rightarrow\mathbb{T}_{\mathbf{t}}^{d,c}$ and the exponential mapping function $\Pi^{exp,c}_{\mathbf{t}}(\mathbf{v}):\mathbb{T}_{\mathbf{t}}^{d,c}\rightarrow\mathbb{L}^{d,c}$ based on the Lorentz model can be written as follows:
\begin{equation}
    \label{ssy0507:exp}
    \left\{
    \begin{aligned}
    & \Pi^{log,c}_{\mathbf{t}}(\mathbf{u})=\sqrt{c} \operatorname{arcosh}\left(-\frac{\langle\mathbf{t}, \mathbf{x}\rangle_{\mathbb{L}}}{c}\right) \frac{\mathbf{u}+\frac{1}{c}\langle\mathbf{t}, \mathbf{u}\rangle_{\mathbb{L}} \mathbf{t}}{\left\|\mathbf{u}+\frac{1}{c}\langle\mathbf{t}, \mathbf{u}\rangle_{\mathbb{L}} \mathbf{t}\right\|_{\mathbb{L}}},\\
    & \Pi^{exp,c}_{\mathbf{t}}(\mathbf{v})=\cosh \left(\|\mathbf{v}\|_{\mathbb{L}}\right) \mathbf{t}+\sinh \left(\|\mathbf{v}\|_{\mathbb{L}}\right) \frac{\mathbf{v}}{\|\mathbf{v}\|_{\mathbb{L}}},
    \end{aligned}
    \right.
\end{equation}
where $\|\mathbf{v}\|_{\mathbb{L}}=\sqrt{\langle\mathbf{v}, \mathbf{v}\rangle_{\mathbb{L}}}$ is the norm of $\mathbf{v}$, $\cosh(\cdot)$ is the Hyperbolic cosine, $\sinh(\cdot)$ is the Hyperbolic sine, and $\operatorname{arcosh}(\cdot)$ is the inverse Hyperbolic cosine. Following the previous work~\cite{DBLP:conf/nips/ChamiYing19}, we set $c$ as a trainable variable. Given the Euclidean embedding $\mathbf{z}_{\mathbb{E}}\in\mathbb{R}^{d-1}$, the corresponding embedding of $\mathbf{z}_{\mathbb{E}}$ in the Hyperbolic space (on the Lorentzian manifold) can be expressed as:
\begin{equation}
    \label{ssy0507:emb_hyper}
    \mathbf{z}_{\mathbb{L}}=\Pi_{\mathbf{o}}^{\exp , c}\left(\left[0, \mathbf{z}_{\mathbb{E}}\right]\right),
\end{equation}
where $\mathbf{o}=\left[\sqrt{c},0,\cdots,0\right]\in\mathbb{L}^{d,c}$ is the origin (satisfying $\langle\mathbf{o}, \mathbf{o}\rangle_{\mathbb{L}}=-c$), $\left[0,\mathbf{z}_{\mathbb{E}}\right]\in\mathbb{T}_{\mathbf{o}}^{d,c}$ is a $d$-dimensional vector in tangent space (satisfying $\langle\left[0,\mathbf{x}_{\mathbb{E}}\right], \mathbf{o}\rangle_{\mathbb{L}}=0$). Using the exponential mapping function in Eq.~\eqref{ssy0507:exp}, $\mathbf{z}_{\mathbb{L}}$ can be further written as
\begin{equation}
    \label{ssy0507:emb_hyper}
    \mathbf{z}_{\mathbb{L}}=\left[\sqrt{c} \cosh \left(\frac{\left\|\mathbf{z}_{\mathbb{E}}\right\|_{2}}{\sqrt{c}}\right), \sqrt{c} \sinh \left(\frac{\left\|\mathbf{z}_{\mathbb{E}}\right\|_{2}}{\sqrt{c}}\right) \frac{\mathbf{z}_{\mathbb{E}}}{\left\|\mathbf{z}_{\mathbb{E}}\right\|_{2}}\right].
\end{equation}
Eq.~\eqref{ssy0507:emb_hyper} shows that we can follow the common protocol to initialize the node embeddings in the Euclidean space, and then transform the Euclidean embeddings into Hyperbolic embeddings by the space mapping. Unless otherwise specified, all embeddings mentioned in the following sections are transformed Hyperbolic embeddings. For brevity, we denote the Hyperbolic embedding $\mathbf{z}_{\mathbb{L}}$ as $\mathbf{z}$.

\vspace{5pt}
\noindent{\textbf{$\bullet$ Knowledge Aggregation}}
Knowledge graphs usually contain many different types of edges (i.e., different types of relationships between nodes), which undoubtedly contain rich semantic information. Inspired by previous work~\cite{DBLP:conf/iclr/VelickovicCucurull18,DBLP:conf/sigir/YangHuang22}, we design a relation-based Lorentzian knowledge aggregation in this section, which is expected to capture rich semantic information in knowledge graphs. Specifically, the message aggregation mechanism between an item $\mathbf{z}_{i}$ and its connected entities $\mathbf{z}_t|_{t\in\mathcal{N}_i}$ can be written as:
\begin{equation}
    \label{ssy0507:message_agg}
    \mathbf{z}_{i} \leftarrow \Pi_{\mathbf{o}}^{exp,c}\left[\Pi_{\mathbf{o}}^{log,c}(\mathbf{z}_i)+\sum_{t\in\mathcal{N}_{i}}\pi(\mathbf{z}_t,\mathbf{z}_i,\mathbf{r}_{ti})\Pi_{\mathbf{o}}^{log,c}(\mathbf{z}_t)\right],
\end{equation}
where $\mathbf{r}_{ti}$ is the embedding of the relation between the entity $t$ and item $i$, and $\pi(\mathbf{z}_t,\mathbf{z}_i,\mathbf{r}_{ti})$ is the normalized attentive weight with $\mathbf{z}_t,\mathbf{z}_i,\mathbf{r}_{ti}$ as input. Specifically, the normalized weight $\pi(\mathbf{z}_t,\mathbf{z}_i,\mathbf{r}_{ti})$ can be expressed as:
\begin{equation}
    \label{ssy0507:norm_weight}
    \pi(\mathbf{z}_t,\mathbf{z}_i,\mathbf{r}_{ti}) = \frac{e^{f\left\{d_{\mathbb{L}}\left[\mathbf{r}_{ti},\mathbf{W}\odot^{c}(\mathbf{z}_{t}\circledast^c\mathbf{z}_{i})\right]\right\}}}{\sum_{t^\prime\in\mathcal{N}_i}e^{f\left\{d_{\mathbb{L}}\left[\mathbf{r}_{t^\prime i},\mathbf{W}\odot^{c}(\mathbf{z}_{t^\prime}\circledast^c\mathbf{z}_{i})\right]\right\}}},
\end{equation}
where $\mathbf{W}$ is a learnable matrix in the tangent space, $\circledast^c$ is the concatenation operation in the Hyperbolic space,
\begin{equation}
    \label{ssy0507:hyper_concat}
    \mathbf{z}_{t} \circledast^c \mathbf{z}_{i}=\Pi_\mathbf{o}^{exp , c}\left(\Pi_\mathbf{o}^{log , c}(\mathbf{z}_{t}) \| \Pi_\mathbf{o}^{log , c}(\mathbf{z}_{i})\right),
\end{equation}
$\odot^c$ is the Lorentzian linear transformation,
\begin{equation}
    \label{ssy0507:hyper_linear}
    \mathbf{W}\odot^{c}(\mathbf{z}_{t}\circledast^c\mathbf{z}_{i})=\Pi_{\mathbf{o}}^{exp,c}\left[\mathbf{W}\Pi_{\mathbf{o}}^{log,c}(\mathbf{z}_{t} \circledast^c \mathbf{z}_{i})\right],
\end{equation}
$d_{\mathbb{L}}$ is the distance function (score function) in the Hyperbolic space to convert a vector to a scalar~\cite{DBLP:conf/iclr/VelickovicCucurull18}:
\begin{equation}
    \label{ssy0507:distance_func}
    d_{\mathbb{L}}\left[\mathbf{r}_{ti}, \mathbf{W}\odot^{c}(\mathbf{z}_{t^\prime}\circledast^c\mathbf{z}_{i})\right]=\\
    \sqrt{c} \operatorname{arcosh}\left[\frac{-\langle\ \mathbf{r}_{ti}, \mathbf{W}\odot^{c}(\mathbf{z}_{t^\prime}\circledast^c\mathbf{z}_{i})\rangle_{\mathbb{L}}}{c}\right],
\end{equation}
and $f\left[d_\mathbb{L}(\mathbf{x}_1,\mathbf{x}_2)\right]$ is the Fermi-Dirac function~\cite{DBLP:journals/PRE/KrioukovPapadopoulos10} (which is widely used to change the monotonicity\footnote{In the Euclidean space, the dot product is often used to convert a vector into a scalar. In the dot product operation, the closer the two vectors $\mathbf{a}_1$ and $\mathbf{a}_2$ are, the larger the value of $\mathbf{a}_1^{\top}\cdot\mathbf{a}_2$ is. However, it is easy to verify that the function $d_{\mathbb{L}}(\cdot)$ has the opposite property, that is, the closer the two vectors $\mathbf{a}_1$ and $\mathbf{a}_2$ are, the smaller the value of $d_{\mathbb{L}}(\mathbf{a}_1,\mathbf{a}_2)$ is. Considering the difference mentioned above, we introduce the Fermi-Dirac function here, which changes the monotonicity by putting $d_{\mathbb{L}}(\cdot)$ on the denominator. Other functions can also be designed to change the monotonicity, but this is beyond the scope of this paper and we leave it for future exploration.}) with hyper-parameters $c_1$, $c_2$:
\begin{equation}
    \label{ssy0507:FermiDirac_function}
    f\left[d_\mathbb{L}(\mathbf{x}_1,\mathbf{x}_2)\right] = \left[{e^{\frac{d_\mathbb{L}(\mathbf{x}_1,\mathbf{x}_2)-c_1}{c_2}}+1}\right]^{-1}.
\end{equation}
The proposed knowledge aggregation mechanism aims to capture both the rich semantic information (relation-aware) and the intrinsic hierarchical structure information (Lorentzian knowledge aggregation) in the graph.
\subsection{Encode the Representations of Users/Items}
After relation-based Lorentzian knowledge aggregation on the knowledge graph $\mathcal{G}_2$ (see Section~\ref{ssy0507:KnowAgg} for details), we then encode the representations of users/items based on the user-item bipartite graph $\mathcal{G}_1$. The basic idea of encoding is to apply the GNN on the user-item bipartite graph $\mathcal{G}_1$, which obtains the representations of the target node (the user/item node) by iteratively aggregating the local information of the target node's multi-hop neighbors. 

In the process of information aggregation, the classic GNN usually includes the steps of feature transformation and non-linear activation. However, recent studies~\cite{DBLP:conf/sigir/HeDeng20,DBLP:conf/www/SunCheng21} have pointed out that feature transformation and non-linear activation are not necessary in recommendation scenarios. This is because feature transformation and non-linear activation can significantly improve the expressive power of the GNN, which often leads to overfitting~\cite{DBLP:conf/sigir/HeDeng20} on the sparse graph data (i.e., the sparse user-item interactions). Furthermore, considering that the mean aggregation operation in the Hyperbolic space is not trivial (due to the Fréchet mean, the generalization of the Euclidean mean in the Hyperbolic space, has no closed-form solution~\cite{DBLP:journals/Fréchet}), we follow previous studies~\cite{DBLP:conf/www/SunCheng21,DBLP:conf/cikm/WangHu21} to perform aggregation in the tangent space. Above all, the aggregation mechanism adopted in this paper can be written as:
\begin{equation}
    \label{ssy0507:encode_user_item}
    \left\{
    \begin{aligned}
    &\mathbf{z}_{u}^{(k+1)}=\Pi_{\mathbf{o}}^{exp,c}\left\{\sum_{i \in \mathcal{N}_{u}} \frac{1}{\sqrt{\left|\mathcal{N}_{u}\right|} \sqrt{\left|\mathcal{N}_{i}\right|}} \Pi_{\mathbf{o}}^{log,c}\left[\mathbf{z}_{i}^{(k)}\right]\right\}, \\
    &\mathbf{z}_{i}^{(k+1)}=\Pi_{\mathbf{o}}^{exp,c}\left\{\sum_{u \in \mathcal{N}_{i}} \frac{1}{\sqrt{\left|\mathcal{N}_{i}\right|} \sqrt{\left|\mathcal{N}_{u}\right|}} \Pi_{\mathbf{o}}^{log,c}\left[\mathbf{z}_{u}^{(k)}\right]\right\},
    \end{aligned}
    \right.
\end{equation}
where $\mathbf{z}_{u}^{(k)}$ is the representation of the user node $u$ at the $k$-th layer of the GNN, $\mathbf{z}_{i}^{(k)}$ is the representation of the item node $u$ at the $k$-th layer of the GNN, $\mathcal{N}_{u}$ is the set of neighbors of the user node $u$, and $\mathcal{N}_{i}$ is the set of neighbors of the item node $i$.
\subsection{Model-augmentation Contrastive Learning}
\label{ssy0507:model_contrast}
Classical contrastive learning needs to perturb the input data to generate different views, which can easily lead to changes in semantics~\cite{DBLP:conf/www/XiaWu22,DBLP:journals/tkde/YuXia23}. In this section, we investigate model-augmentation contrastive learning, which uses model-level augmentations instead of graph structure-level augmentations. For the sake of simplicity, we define an abstract parameterized function containing knowledge aggregation and node representation encoding as $f(\cdot)$, then the process of obtaining the user/item embedding can be mathematically expressed as: $\mathbf{z}_{v} = f(\mathcal{G}_{1},\mathcal{G}_{2},v|\boldsymbol{\Theta}),$ where $\mathbf{z}_{v}$ is the final embedding of user/item node $v$, $\mathcal{G}_{1}$ is a user-item bipartite graph, $\mathcal{G}_{2}$ is a knowledge graph, and $\boldsymbol{\Theta}$ is a tensor of learnable parameters.

\vspace{5pt}
\noindent\textbf{$\bullet$ Augmentations via Dropout}
In the previous method such as KGCL~\cite{DBLP:conf/sigir/YangHuang22}, the augmentation
methods focus on graph structure perturbations (e.g., the interaction graph augmentation and the knowledge graph augmentation). Although the augmentation based on the graph structure perturbation can help improve the performance of the model to some extent, it can easily cause user preference shifting. To address the above problem, we propose to apply Dropout-based augmentation in the knowledge graph-based recommendation,  which avoids directly perturbing the original graph structure. Considering that there are Dropout modules in both knowledge aggregation and representation encoding, we can simply feed the same input twice into the encoder to obtain two different views ($\mathbf{z}_{v}^{D}$ and $\mathbf{z}_{v^{+}}^{D}$) of node $v$, i.e.,
\begin{equation}
    \label{ssy0507:dropout_aug}
    \left\{
    \begin{aligned}
        &\mathbf{z}_{v}^{D} = f(\mathcal{G}_{1},\mathcal{G}_{2},v|\boldsymbol{\Theta},\mathbf{M}_1^{D}),\\
        &\mathbf{z}_{v^{+}}^{D} = f(\mathcal{G}_{1},\mathcal{G}_{2},v|\boldsymbol{\Theta},\mathbf{M}_2^{D}),
    \end{aligned}
    \right.
\end{equation}
where $\mathbf{M}_1^{D}$ is the Dropout mask used in the first forward-passing, $\mathbf{M}_2^{D}$ is the Dropout mask used in the second forward-passing, and the superscript $D$ implies that the augmentation via Dropout is adopted.

\vspace{5pt}
\noindent\textbf{$\bullet$ Augmentations via Cross-Layer Outputs}
The augmentation via cross-layer outputs takes the output of different layers in the convolutional network as different views (equivalent to the model layer dropping), i.e.,
\begin{equation}
    \label{ssy0507:cross_layer_aug}
    \left\{
    \begin{aligned}
        &\mathbf{z}_{v}^{C} = f^{(k_1)}(\mathcal{G}_{1},\mathcal{G}_{2},v|\boldsymbol{\Theta},\mathbf{M}^{D}),\\
        &\mathbf{z}_{v^{+}}^{C} = f^{(k_2)}(\mathcal{G}_{1},\mathcal{G}_{2},v|\boldsymbol{\Theta},\mathbf{M}^{D}),
    \end{aligned}
    \right.  \quad\quad k_1 \neq k_2,
\end{equation}
where $f^{(k_1)}(\cdot)$ is the output of the model at the $k_1$-th layer, $f^{(k_2)}(\cdot)$ is the output of the model at the $k_2$-th layer, the superscript $C$ implies that the augmentation via cross-layer outputs is adopted, and $\mathbf{M}^{D}$ is the dropout mask used in the forward-passing. Note that only one forward-passing is required to get the two different views here, so the dropout masks used in the generation of different views in the Eq.~\eqref{ssy0507:cross_layer_aug} are the same.

\vspace{5pt}
\noindent\textbf{$\bullet$ Augmentations via Model-Pruning}
To ensure the high semantic similarity between the augmented positive sample pair (i.e., avoiding the shift of user preferences), we propose the augmentation via model-pruning in this section. The motivation for applying the model-pruning to augmentation is that the model-pruning aims to discard weights that have little impact on model performance, which is consistent with our goal of avoiding preference changes and ensuring semantic similarity (this is because if the performance of the model after pruning does not change much compared to the performance of the model before pruning, it is reasonable to infer that the semantics of the output of the pruned model and the semantics of the output of the model without pruning are roughly the same). Formally, two different views obtained via the model-pruning can be written as
\begin{equation}
    \label{ssy0507:dropout_aug}
    \left\{
    \begin{aligned}
        &\mathbf{z}_{v}^{P} = f(\mathcal{G}_{1},\mathcal{G}_{2},v|\boldsymbol{\Theta}),\\
        &\mathbf{z}_{v^{+}}^{P} = f(\mathcal{G}_{1},\mathcal{G}_{2},v|\boldsymbol{\Theta}\odot\mathbf{M}_2^{P}),
    \end{aligned}
    \right.
\end{equation}
where the superscript $P$ implies that the augmentation via model-pruning is adopted, $\mathbf{z}_{v}^{P}$ is the output of the model without pruning, $\mathbf{z}_{v^{+}}^{P}$ is the output of the pruned model, and $\mathbf{M}_2^{P}$ is the  pruning mask generated according to the magnitude pruning strategy\footnote{Other more complex pruning strategies can also be applied. Considering that we are only using pruning to generate cotrastive views and do not have strict requirements for the performance of the pruned model, simple magnitude pruning will suffice.} (set the elements in $\boldsymbol{\Theta}$ greater than the $\xi$ to 0),
\begin{equation}
    \label{ssy0507:prune_mask}
    \mathbf{M}_2^{P} = \mathbbm{1}(|\boldsymbol{\Theta}|>\xi),
\end{equation}
with the indicator function $\mathbbm{1}(\cdot)$ and the threshold $\xi$ that determined by the pruning ratio. It should be noted that in order to evaluate the effectiveness of the pruning-based augmentation proposed in this part, we disable all Dropout modules in the model here (see Section \ref{ssy0507:experiment} for more details). 

\subsection{Training and Optimization}
\textbf{$\bullet$ Recommendation Loss}
Given the final Hyperbolic representations of users ($\mathbf{z}_{u}|u\in\mathcal{U}$) and items ($\mathbf{z}_{i}|i\in\mathcal{I}$), the recommendation loss can be written as:
\begin{equation}
    \label{ssy0507:rec_loss}
    \mathcal{L}_{1}=\sum_{u \in \mathcal{U}} \sum_{i \in \mathcal{N}_{u}} \sum_{i^{\prime} \notin \mathcal{N}_{u}}-\log \sigma\left(\hat{y}_{u, i}-\hat{y}_{u, i^{\prime}}\right),
\end{equation}
where $i$ is an item that the user $u$ interacts with, and $i^\prime$ is an item that user $u$ does not interact with, $\sigma$ is the Sigmoid function, and $\hat{y}_{u, i}$ is used to measure user $u$'s preference on item $i$,
\begin{equation}
    \hat{y}_{u, i} = \Pi_{\mathbf{o}}^{log,c}\left(\mathbf{z}_{u}\right)\Pi_{\mathbf{o}}^{log,c}\left(\mathbf{z}_{i}\right).
\end{equation}
\textbf{$\bullet$ Contrastive Loss}
The goal of contrastive learning is to learn representations so that representations of similar pairs are close to each other and representations of dissimilar pairs are far apart. The contrastive loss can be written as:
\begin{equation}
    \label{ssy0507:contrast_loss}
        \mathcal{L}_{c}^{\mathcal{V}}=\sum_{i \in \mathcal{V}}-\log \frac{\exp \left[{\Pi_{\mathbf{o}}^{log,c}\left(\mathbf{z}_{i}^{Aug}\right)\Pi_{\mathbf{o}}^{log,c}\left(\mathbf{z}_{i^{+}}^{Aug}\right)}/{\tau}\right]}{\sum\limits_{j \in \{i^{+}\}\cup\mathcal{V}\backslash\{i\}} \exp \left[{\Pi_{\mathbf{o}}^{log,c}\left(\mathbf{z}_{i}^{Aug}\right)\Pi_{\mathbf{o}}^{log,c}\left(\mathbf{z}_{j}^{Aug}\right)}/{\tau}\right]},
\end{equation}
where $\mathbf{z}_{i}^{Aug}$ and $\mathbf{z}_{i^{+}}^{Aug}$ are two views generated by the augmentation strategy ``\emph{Aug}'' ($\emph{Aug}\in\{D,C,P\}$ is one of the augmentation methods proposed in Section \ref{ssy0507:model_contrast}), $\tau$ is the temperature hyper-parameter, and $\mathcal{L}_{c}^{\mathcal{V}}$ is the contrastive loss computed on the set $\mathcal{V}$. It should be noted that we calculate the contrastive loss on the set of users ($\mathcal{L}_{c}^{\mathcal{U}}$) and the set of items ($\mathcal{L}_{c}^{\mathcal{I}}$), respectively, and then add them up as the final contrastive loss, i.e., $\mathcal{L}_{2} = \mathcal{L}_{c}^{\mathcal{U}}+\mathcal{L}_{c}^{\mathcal{I}}$.

\vspace{5pt}
\noindent\textbf{$\bullet$ Joint Training}
We adopt a multi-task training strategy where the auxiliary contrastive learning task and the main recommendation task are
jointly optimized. Specifically, we jointly optimize the recommendation loss and the contrastive loss, i.e.,
\begin{equation}
    \label{ssy0507:tol_loss}
    \mathcal{L} = \mathcal{L}_{1} + \lambda\mathcal{L}_{2},
\end{equation}
where $\lambda$ is a hyper-parameter to balance the effects of different losses.

\section{Experiments}
\label{ssy0507:experiment}
Extensive experiments are conducted to evaluate the performance of the proposed methods. Our experiments aim to answer the following research questions (RQs):
\begin{itemize}
\item[$\bullet$]  \textbf{RQ1:} Can the proposed methods achieve performance improvements compared to their counterparts? 
\item[$\bullet$] \textbf{RQ2:} What is the impact of different components in the proposed methods on the overall performance?
\item[$\bullet$] \textbf{RQ3:} How do the proposed methods perform under different hyper-parameter settings and how about their time efficiency?
\end{itemize}

\subsection{Experiment Settings}
\begin{itemize}
\item{\textbf{Datasets}
We conduct experiments on three benchmark datasets: \emph{Yelp2018}\footnote{https://www.yelp.com/dataset}, \emph{Amazon-Book}\footnote{http://jmcauley.ucsd.edu/data/amazon/}, and \emph{MovieLens-20M}\footnote{https://grouplens.org/datasets/movielens/}. The \emph{Yelp2018} is a widely used dataset for business recommendations, which is collected by Yelp, a well-known platform for recommending restaurants. The transaction records after Jan. 1st, 2018 are used in the experiments. The \emph{Amazon-Book} is a widely adopted dataset for product recommendations, which is collected from Amazon, one of the largest online shopping platforms in the world. The \emph{MovieLens-20M} is a dataset for movie recommendations, which is collected from the MovieLens website (containing approximately 20 million explicit ratings). The statistics of the above three datasets are provided in Table \ref{ssy0507:dataset_detail}. The split of the datasets follows previous studies~\cite{DBLP:conf/sigir/YangHuang22,DBLP:conf/kdd/WangHe19}. For each dataset, we randomly select 80\% of the interaction records of each user to constitute the training set, and treat the remaining as the test set. From the training set, 10\% of interaction records are randomly selected as the validation set to tune hyper-parameters.}

\begin{table}[htbp]
  \centering
  \caption{Statistics of datasets.}    
  \begin{tabular}{cccc}
    \toprule[1.5pt]
    dataset & Yelp2018 & Amazon-Book & MovieLens-20M \\
    \midrule
    \#Users         & 45,919    & 70,679    & 138,159 \\
    \#Item          & 45,538    & 24,915    & 16,953 \\
    \#Interactions  & 1,183,610 & 846,434   & 13,501,622 \\
    \#Entities      & 47,472    & 29,714    & 85,615 \\
    \#Relations     & 42        & 39        & 32 \\
    \#Triplets      & 869,603   & 686,516   & 499,474 \\
    \bottomrule[1.5pt]
    \end{tabular}
  \label{ssy0507:dataset_detail}
\end{table}

\vspace{5pt}
\item{\textbf{Metrics and Baselines} 
In the experiments, we used $\emph{Recall@K}$ $(R@K)$ and $\emph{NDCG@K}$ $(N@K)$ for overall evaluation, where $K\in\{10,20\}$. All-ranking strategy is adopted in the experiments, which is consistent with the settings of previous studies~\cite{DBLP:conf/kdd/WangHe19,DBLP:conf/sigir/YangHuang22}. Specifically, we regard all uninteracted items of the user $u$ as negative samples when inferring user $u$’s preference. In order to verify the effectiveness of the proposed methods, we compare the proposed methods with the following baselines:
\vspace{2pt}
\begin{itemize}[leftmargin=*]
    \item {\textbf{BPR}~\cite{arxiv:RendleFreudenthaler12}. The BPR is a classic recommender model that ranks candidate items with a pairwise ranking loss.}
    \item {\textbf{NCF}~\cite{DBLP:conf/www/HeLiao17}. The NCF is the first attempt to learn the user-item interaction function by using a multi-layer perception, which replaces the inner product in the matrix factorization model with neural networks.}
    \item {\textbf{LightGCN}~\cite{DBLP:conf/sigir/HeDeng20}. The LightGCN is a GCN-based recommender model, which simplifies the convolution operation on the user-item interaction graph by discarding feature transformations and nonlinear activations.}
    \item {\textbf{SGL}~\cite{DBLP:conf/sigir/WuWang21}. The SGL is a method based on the contrastive learning framework. The core idea of the SGL is to leverage self-supervised signals to assist the traditional supervised recommendation task.}
    \item {\textbf{CKE}~\cite{DBLP:conf/kdd/ZhangJing16}. The CKE is a regularization-based method, which aims to utilize the heterogeneous information in the knowledge base to improve the performance of the recommender system.}
    \item {\textbf{KGCN}~\cite{DBLP:conf/www/WangZhao19}. The KGCN aims to capture high-order structural information and semantic information in the knowledge graph, which is beneficial for modeling users' long-distance interests.} 
    \item {\textbf{CFKG}~\cite{DBLP:journals/alg/AiAzizi18}. The CFKG aims to incorporate both user behaviors and external knowledge of items by constructing a user-item knowledge graph.}
    \item {\textbf{RippleNet}~\cite{DBLP:conf/cikm/WangZhang18}. The RippleNet is a propagation-based recommender model. The RippleNet automatically propagates users' potential preferences by using a memory-like network.}
    \item {\textbf{KGNN-LS}~\cite{DBLP:conf/kdd/WangZhang19}. The KGNN-LS introduces the label smoothness regularization into knowledge-aware graph neural networks. The introduction of the label smoothness regularization improves the robustness of the model.}
    \item {\textbf{KGAT}~\cite{DBLP:conf/kdd/WangHe19}. The KGAT is a representative regularization-based recommender model, which aims to capture high-order connectivities in the collaborative knowledge graph by using the well-designed attentive embedding propagation layer.}
    \item {\textbf{MCCLK}~\cite{DBLP:conf/sigir/ZouWei22}. The core idea of MCCLK is to perform multi-level contrastive learning among local and global views.}
    \item {\textbf{KGCL}~\cite{DBLP:conf/sigir/YangHuang22}. The KGCL is a contrastive-based recommender model, which uses structural-level knowledge graph augmentations to guide the contrastive learning paradigm.}
\end{itemize}
}
\vspace{5pt}
\item{\textbf{Implementation Details}
{For fair comparisons, we set the embedding dimension as 64 and the batch size as 2,048 for all models. For all the baselines, we follow the official hyper-parameter settings from the original papers. The hyper-parameters $c_1$ and $c_2$ in Eq.~\eqref{ssy0507:FermiDirac_function} are searched in $\{0.5,1.0,1.5\}$. The temperature $\tau$ in Eq.~\eqref{ssy0507:contrast_loss} is tuned in $\{0.2,0.4,0.6,0.8,1.0\}$. The hyper-parameter $\lambda$ in Eq.~\eqref{ssy0507:tol_loss} and the dropout/pruning ratio used in the augmentation are tuned in $\{0.1,$ $0.2,$ $0.3,$ $0.4,$ $0.5\}$.} The number of layers is set as 3 in the process of encoding user/item representations. For the augmentation via cross-layer outputs, we take the output at different layers as different views to tune the model. The proposed methods are optimized by the Adam.}
\end{itemize}

\begin{table}
  \centering
  \caption{Performance of all methods on three public datasets. The best results are marked in bold and the second-best results are underlined.}
    \scalebox{0.71}{
    \begin{tabular}{c|cccc|cccc|cccc}
    \toprule[1.5pt]
    Dataset & \multicolumn{4}{c|}{Yelp2018}  & \multicolumn{4}{c|}{Amazon-Book} & \multicolumn{4}{c}{MovieLens-20M}  \\
    \cmidrule{1-5} \cmidrule{6-9}  \cmidrule{10-13}     
    Metric & {N@10} & {R@10} & {N@20} & {R@20} & {N@10} & {R@10} & {N@20} & {R@20} & {N@10} & {R@10} & {N@20} & {R@20} \\
    \midrule
    BPR   & 0.0261 & 0.0315 & 0.0336 & 0.0521 & 0.0473 & 0.0746 & 0.0597 & 0.1152 & 0.2438 & 0.226 & 0.2684 & 0.3313 \\
    NCF   & 0.022 & 0.0263 & 0.0292 & 0.0464 & 0.0431 & 0.0692 & 0.0542 & 0.1053 & 0.1985 & 0.2046 & 0.2306 & 0.3093 \\
    LightGCN & 0.033 & 0.0399 & 0.0443 & 0.0682 & 0.0573 & 0.0895 & 0.0736 & 0.1398 & 0.2591 & 0.2333 & 0.2811 & 0.3383 \\
    SGL   & 0.0386 & 0.0453 & 0.0475 & 0.0719 & 0.0652 & 0.1018 & 0.0766 & 0.1445 & 0.2385 & 0.2129 & 0.2592 & 0.3099 \\
    CKE   & 0.0342 & 0.0397 & 0.0428 & 0.0655 & 0.0548 & 0.0854 & 0.0676 & 0.1271 & 0.2442 & 0.2392 & 0.2689 & 0.3453 \\
    KGCN  & 0.024 & 0.0307 & 0.0315 & 0.0528 & 0.0391 & 0.0693 & 0.0506 & 0.1109 & 0.2166 & 0.2202 & 0.2441 & 0.3252 \\
    CFKG  & 0.033 & 0.0382 & 0.0415 & 0.0638 & 0.0618 & 0.0929 & 0.0761 & 0.1397 & 0.2044 & 0.2171 & 0.2349 & 0.3227 \\
    RippleNet & 0.019 & 0.0249 & 0.025 & 0.0425 & 0.0381 & 0.067 & 0.0489 & 0.1059 & 0.2035 & 0.2028 & 0.228 & 0.302 \\
    KGNN-LS & 0.0263 & 0.0331 & 0.0338 & 0.0555 & 0.0393 & 0.0681 & 0.0507 & 0.1096 & 0.2166 & 0.2202 & 0.244 & 0.3252 \\
    KGAT  & 0.0339 & 0.0418 & 0.0434 & 0.0704 & 0.0561 & 0.0846 & 0.0706 & 0.1323 & 0.2428 & 0.2369 & 0.2678 & 0.3449 \\
    MCCLK  & 0.0366 & 0.043 & 0.0463 & 0.0723 & 0.0644 & 0.0954 & 0.0788 & 0.1424 & 0.2450 & 0.2211 & 0.2674 & 0.3244 \\
    KGCL  & 0.0386 & 0.0455 & 0.0493 & 0.0756 & 0.0643 & 0.0999 & 0.0793 & 0.1496 & 0.2527 & 0.2342 & 0.2769 & 0.3392 \\
    \midrule
    HCMKR-P & 0.039 & 0.0457 & 0.0497 & 0.0758 & 0.0661 & 0.1023 & 0.081 & 0.1512 & 0.2576 & 0.234 & 0.2803 & 0.3378 \\ 
     Improve & 1.04\%$\uparrow$ & 0.44\%$\uparrow$ & 0.81\%$\uparrow$ & 0.26\%$\uparrow$ & 1.38\%$\uparrow$ & 0.49\%$\uparrow$ & 2.14\%$\uparrow$ & 1.07\%$\uparrow$ & N/A   & N/A   & N/A   & N/A \\
    \midrule
    HCMKR-D & \underline {0.0395} & \underline{0.0468} & \underline{0.0526} & \underline{0.0809} & \underline{0.0663} & \underline{0.1038} & \underline{0.0820} & \underline{0.1548} & \underline{0.2862} & \underline{0.2573} & \underline{0.3086} & \underline{0.3673} \\
     Improve & \underline{2.33\%$\uparrow$} & \underline{2.86\%$\uparrow$} & \underline{6.69\%$\uparrow$} & \underline{7.01\%}$\uparrow$ & \underline{1.69\%$\uparrow$} & \underline{1.96\%$\uparrow$} & \underline{3.40\%$\uparrow$} & \underline{3.48\%$\uparrow$} & \underline{10.46\%$\uparrow$} & \underline{7.57\%$\uparrow$} & \underline{9.78\%$\uparrow$} & \underline{6.37\%$\uparrow$} \\
    \midrule
    HCMKR-C & \textbf{0.0395} & \textbf{0.0469} & \textbf{0.0542} & \textbf{0.0835} & \textbf{0.0703} & \textbf{0.1099} & \textbf{0.0869} & \textbf{0.1622} & \textbf{0.2875} & \textbf{0.2608} & \textbf{0.3121} & \textbf{0.3739} \\
     Improve & \textbf{2.33\%$\uparrow$} & \textbf{3.08\%$\uparrow$} & \textbf{9.94\%$\uparrow$} & \textbf{10.45\%$\uparrow$} & \textbf{7.82\%$\uparrow$} & \textbf{7.96\%$\uparrow$} & \textbf{9.58\%$\uparrow$} & \textbf{8.42\%$\uparrow$} & \textbf{10.96\%$\uparrow$} & \textbf{9.03\%$\uparrow$} & \textbf{11.03\%$\uparrow$} & \textbf{8.28\%$\uparrow$} \\
    \bottomrule[1.5pt]
    \end{tabular}
\label{ssy0507:overall_performance}}
\end{table}
\subsection{Performance Comparison (RQ1)}
\label{ssy0507:perform_compare}
In this section, we compare the proposed HCMKR to its counterparts by conducting experiments on three public datasets. For the sake of brevity, we use the abbreviations HCMKR-P, HCMKR-D, and HCMKR-C to represent the pruning-based, Dropout-based, and the cross-layer-based method, respectively. 

We first evaluate the proposed methods on the Top-K recommendation and provide the experimental results in Table~\ref{ssy0507:overall_performance}. The proposed methods demonstrate superiority over baselines in most cases. As shown in Table~\ref{ssy0507:overall_performance}, the HCMKR-C always ranks first and achieves a performance gain of 2.33\% to 11.03\% compared to the best-performing baseline. The HCMKR-D is the second-best method, which achieves a performance improvement of 1.69\% to 10.46\% compared to the best-performing baseline. The HCMKR-P also performs well in the Top-K recommendation, which achieves performance gains in most cases. Note that in order to evaluate the effectiveness of the proposed pruning-based augmentation, the Dropout module in the HCMKR-P is turned off (since the Dropout module itself can be used for augmentation). The closure of the Dropout module affects the overall performance of the HCMKR-P to a certain extent, which leads to the result that the performance of the HCMKR-P is not as good as the other two proposed methods. Although turning off the Dropout module affects the performance of the HCMKR-P, the proposed HCMKR-P is still comparable to the best baseline method KGCL. The introduction of Hyperbolic geometry and model-level augmentations are two thrusts for the performance improvement of the proposed methods. Specifically, the introduction of Hyperbolic geometry enables the model to capture the intrinsic hierarchical structures of graphs and the model-level augmentation avoids user preferences shift between the augmented positive pair. 

Overall, the KGCL is the best-performing baseline in Table~\ref{ssy0507:overall_performance}. The good performance of the KGCL can be attributed to its well-designed knowledge graph contrastive learning framework, which aims to learn high-quality representations by introducing self-supervised signals (obtained from the knowledge graph augmentation process). However, there are two shortcomings in KGCL. First, KGCL focuses on structural-level augmentations, which makes it difficult to maintain user interest context and avoid preference
shifts. Second, KGCL is tailored to Euclidean space, which makes it lack the ability to capture the intrinsic hierarchical structures of graphs. The above two shortcomings lead to the performance gap between the KGCL and our methods.

\begin{figure}
    \centering
    \includegraphics[scale=0.35]{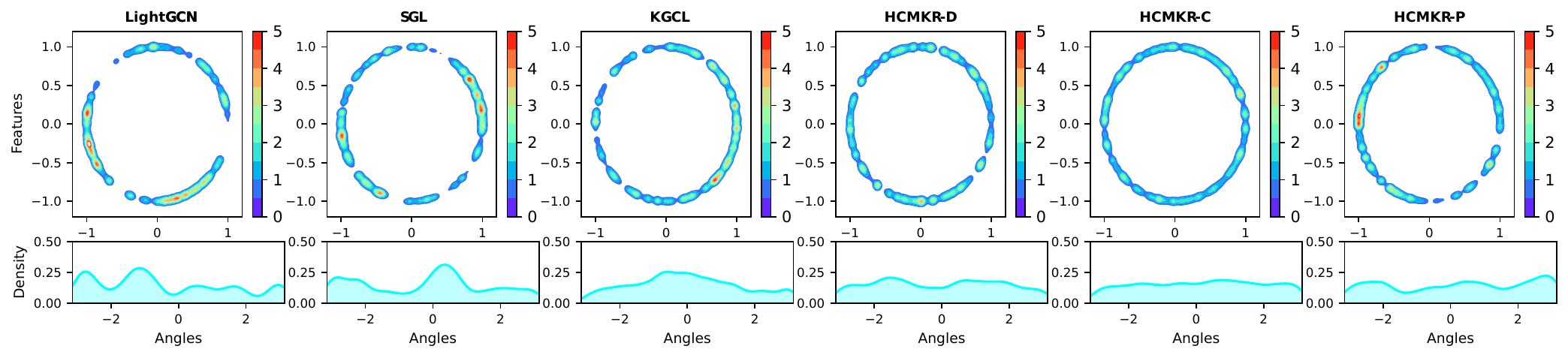}
    \caption{Visualizing the distribution of embeddings (Yelp2018).}
    \label{ssy0507:kde-yelp}
\end{figure}

\begin{figure}
    \centering
    \includegraphics[scale=0.35]{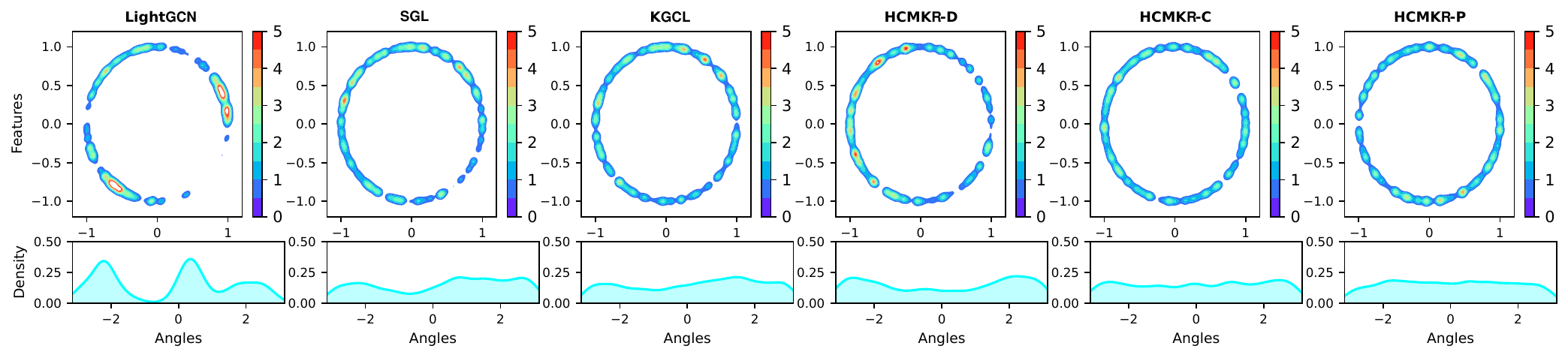}
    \caption{Visualizing the distribution of embeddings (Amazon-Book).}
    \label{ssy0507:kde-amazon}
\end{figure}

To further demonstrate the benefits of the proposed methods, we then plot embedding distributions with Gaussian kernel density estimation (as shown in Fig.~\ref{ssy0507:kde-yelp} and Fig.~\ref{ssy0507:kde-amazon}, we first use TSNE to map the embeddings to a 2D space, and then visualize their Gaussian kernel density estimations). Existing work has shown that contrastive learning is closely related to the uniformity of learned representations (i.e., the distribution that preserves the maximum information of the learned representations). As shown in Fig.~\ref{ssy0507:kde-yelp} and Fig.~\ref{ssy0507:kde-amazon}, the representations learned by the proposed methods are relatively uniformly distributed in the space. For example, the distribution of the learned representations of the HCMKR-C is more uniform than that of its counterparts and its density-angle curve is smoother than that of other baselines. 

\begin{figure}[htbp]
  \centering
  \vfill
    \centering
    \subfigure[HCMKR-C.]{
		\includegraphics[scale=0.25]{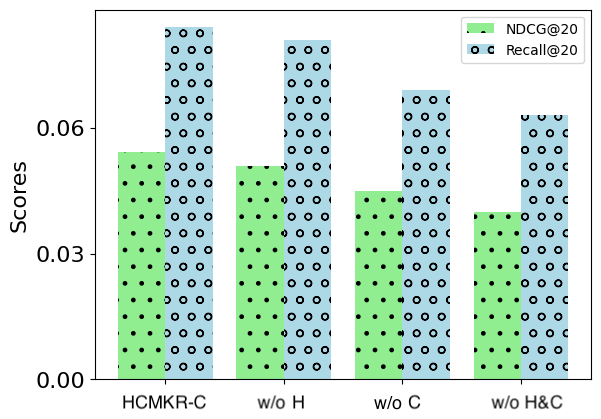}}
    \subfigure[HCMKR-P.]{
		\includegraphics[scale=0.25]{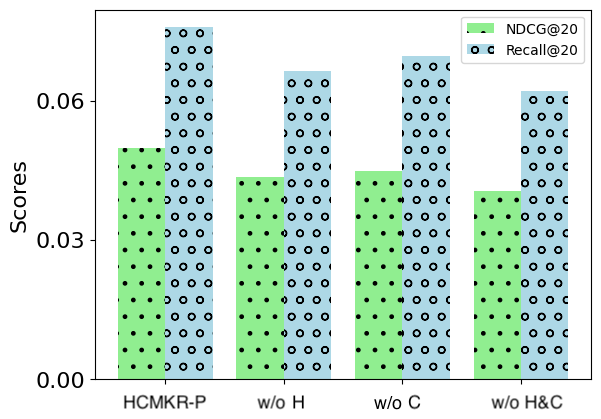}}
    \subfigure[HCMKR-D.]{
		\includegraphics[scale=0.25]{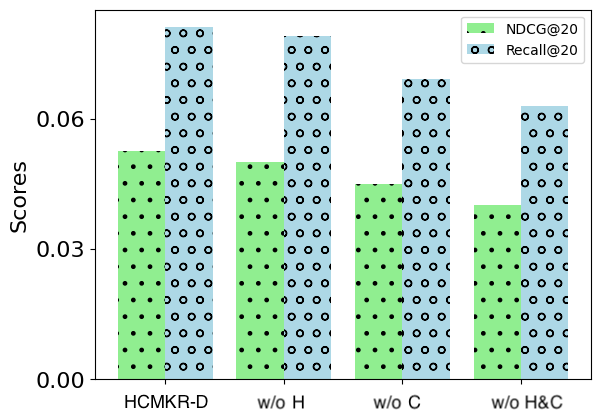}}
    \caption{Ablation experiments of the proposed methods.}
    \label{ssy0507:aba-experiment}
\end{figure}
\subsection{Ablation Study (RQ2)}
For the sake of brevity, we use the abbreviations ``C'' and ``H'' to denote ``using the model-augmentation contrastive learning'' and ``using the Hyperbolic geometry'', respectively. The experimental results based on Yelp2018 are presented in  Fig.~\ref{ssy0507:aba-experiment}. In experiments without using the Hyperbolic geometry and the model-augmentation contrastive learning, we use the Euclidean version of the relation-aware knowledge aggregation proposed in~\cite{DBLP:conf/sigir/YangHuang22} and use the Euclidean version of Eq.~\eqref{ssy0507:encode_user_item} to encode the representations of users/items. It can be seen from Fig.~\ref{ssy0507:aba-experiment} that the complete method always outperforms its variants, that is, the best performance is achieved by using both the Hyperbolic geometry and the model-augmentation contrastive learning. We also observe that the performance of HCMKR-P's variants is somewhat inferior to that of the other two proposed methods' variants, mainly due to the deactivation of the Dropout module.

\begin{figure}
  \centering
  \vfill
  \begin{minipage}{0.5\textwidth}
    \centering
    \subfigure[Drop ratio.]{
		\includegraphics[width=1.13in]{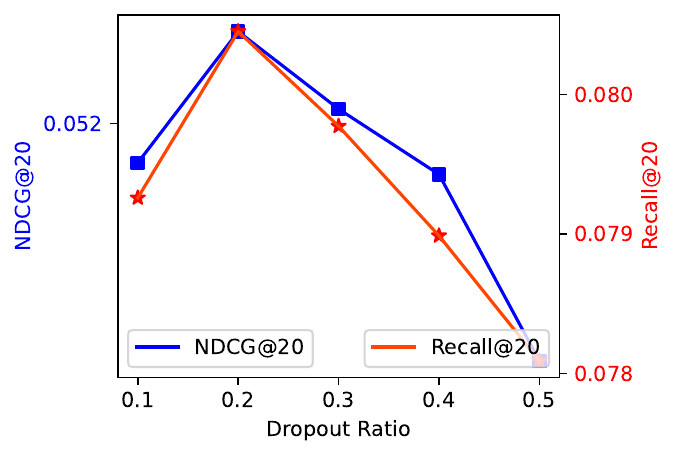}}\hspace{-1em}
    \subfigure[Prune ratio.]{
		\includegraphics[width=1.13in]{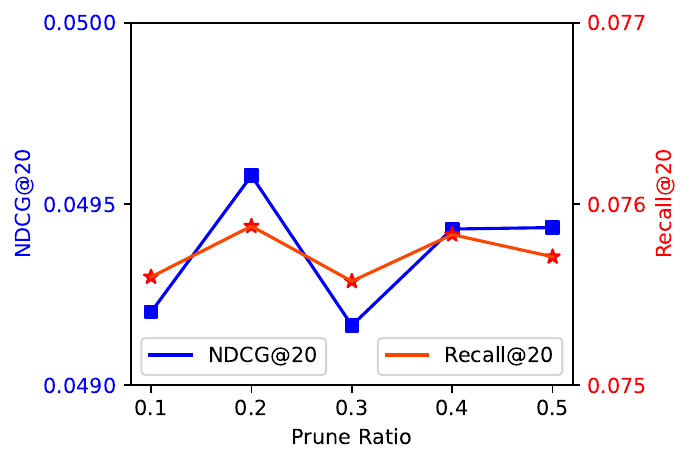}}\\
    \subfigure[Layer.]{
		\includegraphics[width=1.13in]{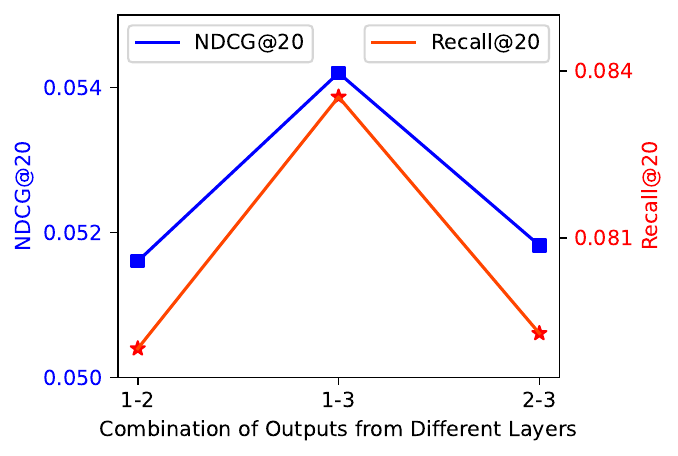}}\hspace{-1em}
    \subfigure[$\lambda$.]{
		\includegraphics[width=1.0in]{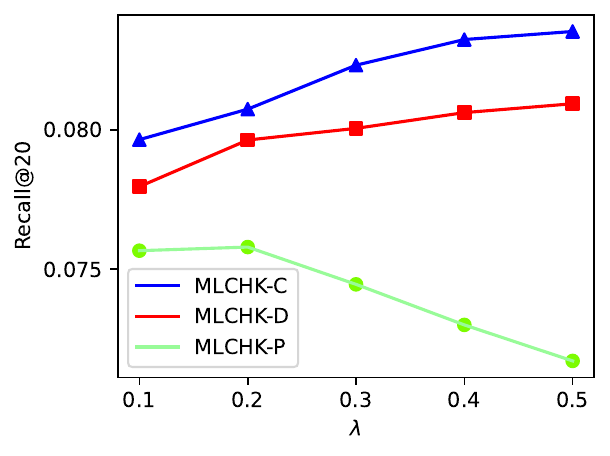}}
    \caption{Parameter sensitivity of Dropout ratio, prune ratio, combination of different layers, and $\lambda$ (Yelp2018).}
    \label{ssy0507:para_sensitivity}
  \end{minipage}%
  \begin{minipage}{0.5\textwidth}
    \centering
    \tabcolsep=0.2cm
    \captionof{table}{Training time cost per epoch.}
    \scalebox{1}{
    \begin{tabular}{ccc}
    \toprule[1.5pt]
    Model & Yelp2018 & Amazon-Book \\
    \midrule
    KGCL  & 323.2s & 147.1s \\
    MCCLK & 1036.7s & 542.3s \\
    \midrule
    \addlinespace[1ex]
    HCMKR-P & 413.6s & 185.2s \\
    \emph{time cmp.} & \emph{$\times$1.28} & \emph{$\times$1.26}\\
    \addlinespace[0.8ex]
    HCMKR-D & 412.2s & 186.3s \\
    \emph{time cmp.} & \emph{$\times$1.28} & \emph{$\times$1.27}\\
    \addlinespace[0.8ex]
    HCMKR-C & 283.1s & 129.7s \\
    \emph{time cmp.} & \textbf{\emph{$\times$0.88}} & \textbf{\emph{$\times$0.88}}\\
    \bottomrule[1.5pt]
    \end{tabular}
    }
    \label{ssy1210:timecmp}
    \vfill 
  \end{minipage}
\end{figure}

\subsection{Parameter Sensitivity and Time Efficiency(RQ3)}
Fig.~\ref{ssy0507:para_sensitivity}(a) shows that a large Dropout ratio is detrimental to the performance of the model. This is possibly because the large Dropout ratio makes the two views generated for contrastive learning too dissimilar, thus leading to the result that the model is guided to train in an unreasonable direction. A similar phenomenon can also be seen in Fig.~\ref{ssy0507:para_sensitivity}(b), i.e., a large prune ratio will cause performance degradation. In Fig.~\ref{ssy0507:para_sensitivity}(c), we investigate the impact of combining the outputs of different
layers. For the sake of brevity, we use the abbreviation ``$i$-$j$'' to denote using the output of the $i$-th layer and the output of the $j$-th layer as two different views in contrastive learning. As shown in Fig.~\ref{ssy0507:para_sensitivity}(c), the model achieves the best performance on Yelp2018 when using the combination ``1-3''. We also plot the impact of the balancing coefficient $\lambda$ based on the Yelp2018 in Fig.~\ref{ssy0507:para_sensitivity}(d), and the results show that the HCMKR-C, the HCMKR-D, and the HCMKR-P achieve the best performance when $\lambda=0.5$, $\lambda=0.5$ and $\lambda=0.2$, respectively. As shown in Table~\ref{ssy1210:timecmp}, compared to the state-of-the-art contrastive learning-based knowledge-aware methods, the proposed method has comparable running time (even faster in some cases) to KGCL and significantly outperforms MCCLK. In other words, the introduction of hyperbolic computation does not significantly increase computational overhead.

\section{Related Work}
\textbf{Knowledge-aware Recommendation} Knowledge-aware recommender systems have recently drawn great attention from researchers due to their potential in introducing side information for items. According to how the information in the knowledge graph is used, existing methods can be divided into the path-based method~\cite{DBLP:conf/kdd/HuShi18}, the embedding-based method~\cite{DBLP:conf/kdd/ZhangJing16,DBLP:conf/www/HeLiao17,DBLP:journals/PR/SunLiu23,DBLP:conf/aaai/MaMa21}, and the GNN-based method~\cite{DBLP:conf/kdd/WangHe19,DBLP:conf/sigir/YangHuang22,DBLP:conf/sigir/ZouWei22,DBLP:conf/wsdm/XuanLiu23,arxiv:SunRen24}. In recent years, inspired by the research on contrastive learning in the fields of computer vision and natural language processing, a lot of work has been carried out to study the application of contrastive learning in recommender systems~\cite{DBLP:conf/wsdm/QiuHuang22,DBLP:conf/tkde/WeiHu24,DBLP:conf/wsdm/ChenHuang23}. The KGCL~\cite{DBLP:conf/sigir/YangHuang22} is one of the most famous contrastive learning-based methods designed for the knowledge-aware recommender system, which uses structural-level knowledge graph augmentations to guide the contrastive learning paradigm. In this paper, contrastive learning serves as a powerful tool to help the model learn high-quality representations.

\vspace{5pt}
\noindent\textbf{Hyperbolic Representation Learning} Representation learning plays an important role in the field of machine learning \cite{DBLP:conf/sigir/YangHuang22,DBLP:conf/eusipco/SunLiu20,DBLP:journals/SP/SunLiu21}, which is mainly used to automatically extract useful information from data. Representation learning is usually performed in the Euclidean space. However, representation learning in the Euclidean space cannot effectively capture the underlying hierarchical structure contained in unstructured data (e.g., graph data)~\cite{DBLP:conf/icml/NickelKiela18}. To solve this problem, Maximillian Nickel and Douwe Kiela proposed to perform representation learning in the Hyperbolic space~\cite{DBLP:conf/nips/NickelKiela17}. Inspired by the above work, Hyperbolic representation learning has recently been widely applied to many graph-based data mining tasks, such as node classification~\cite{DBLP:conf/nips/ChamiYing19}, and link prediction~\cite{DBLP:conf/www/SunCheng21}.

\section{Conclusion}\label{sec:conclusion}
In this paper, we propose Hyperbolic contrastive learning with model-augmentation for knowledge-aware recommendation. First, in order to capture the underlying hierarchical structure contained in the graph data, we design a Lorentzian knowledge aggregation mechanism and encode the representations of users/items in the Hyperbolic space. Second, we propose three model-level augmentation techniques (i.e., the augmentation via Dropout, cross-layer outputs, and model-pruning), which hopefully maintain user interest context and avoid preference
shift. Extensive experiments on three benchmark datasets verify the superiority of the proposed methods over various state-of-the-art baselines.



%
%
%
%

\end{document}